\documentclass[intlimits,twoside,a4paper]{article}
\usepackage{amsmath,amssymb}
\usepackage{graphicx}
\usepackage[T2A]{fontenc}
\usepackage[cp1251]{inputenc}

\usepackage{cmpj}

\issue{2011}{14}{2}{23004}

\doinumber{10.5488/CMP.14.23004}

\newcommand{\kB}{k_{\rm B}}
\newcommand{\rf}{\rm ref}

\newcommand{\YY}{\rm Y}
\newcommand{\ep}{{\rm EXP6}}

\title[2-Yukawa and EXP6 fluids]%
{Virial coefficients and vapor-liquid equilibria of the EXP6 and 2-Yukawa fluids}
\author[J. Krej\v{c}\'{\i} \textsl{et al.}]
{J. Krej\v{c}\'{\i}\refaddr{label1},
        I. Nezbeda\refaddr{label1,label2},
        R. Melnyk\refaddr{label3},
        A. Trokhymchuk\refaddr{label3}
}
\addresses{
\addr{label1} Faculty of Science, J. E. Purkinje University,
 400 96 \'{U}st\'{\i} nad Labem, Czech Republic
\addr{label2} Institute of Chemical Process Fundamentals, Academy
of Sciences, 165 02 Prague 6, Czech Republic \addr{label3}
Institute for Condensed Matter Physics
of the National Academy of Sciences of Ukraine,\\
1 Svientsitskii Str., 79011 Lviv, Ukraine}

\date{Received 26 April 2011, in final form 11 May 2011}

\authorcopyright{J.~Krej\v{c}\'{\i}, I.~Nezbeda, R.~Melnyk, A.~Trokhymchuk, 2011}

\begin{document}

\maketitle

\begin{abstract}
Virial coefficients $B_2$ through $B_4$ and the vapor-liquid
equilibria for the EXP6 and 2-Yukawa (2Y) fluids have been
determined using numerical integrations and Gibbs ensemble
simulations, respectively. The chosen 2Y models have been recently
determined as an appropriate reference fluid for the considered
EXP6 models.

\pacs 05.20, 51.30
\end{abstract}

\section{Introduction}

Simple fluids, i.e., the fluids whose molecules interact via a
spherically symmetric potential, $u=u(r)$, are most commonly
modeled as Lennard-Jones (LJ) fluids. When applied to simple real
fluids, the LJ performs reasonably well at subcritical and
slightly supercritical conditions. However, for quite obvious
reasons it fails at high temperatures/pressures: Repulsive
interactions at these conditions are very soft and it has been
well established that the EXP6 potential provides much more
faithful description of the intermolecular
interactions~\cite{WhyEXP6}. Furthermore, the exponential
repulsion agrees with molecular beam scattering data as is also
known from theoretical studies. It is therefore highly desirable,
particularly from the point of geochemical and industrial
(conditions of detonations and propagation of shock waves)
applications, to reach a similar level of understanding and
theoretical description of the EXP6 fluids similar to that of the
LJ fluid.

A number of simulation data for the EXP6 fluid are available in
literature along with early theoretical attempts
(see~\cite{KNMT_2Y} and references therein). The problem of theory
 is that all common methods are based on the assumption of the
presence of very steep repulsions at short separations and thus
make use of, either directly or indirectly, the known properties
of the fluid of hard spheres. Consequently, they are not
applicable to models with soft repulsion. To overcome this problem
we have recently developed an alternative to HS-based theories, a
theory based on the Yukawa  (Y) model as a
reference~\cite{Yreview}. The Y potential seems to be a
`universal' simple fluid model because it is possible, by changing
its parameter [see equation~(\ref{Ypot}) below], to change both
its range and repulsive softness. This is the reason why the Y
potential is used in applications to describe of a variety of
physical phenomena (see~\cite{Yappl} and references therein).

The Y potential is a model belonging to the family of van der
Waals models, i.e., the model with a hard core and approximating
interactions outside the core. A large body of results, both
theoretical and simulation, is available in literature (see, e.g.,
references~\cite{Yref1,Yref2,Yref3} and references therein) for
the Y fluid. However, to apply the Y potential to more realistic
models with a soft repulsive part (i.e., without a hard core), it
is necessary to use a combination of two (or even more) Y
functions which results in a model without any hard core.

In a recent paper~\cite{KNMT_2Y} we investigated, by means of
molecular simulations, the structure of the EXP6 fluids and
formulated the criteria for determining a Y model [more accurately,
two Yukawa model (2Y)] which could be used as a reference system
for developing an analytic theory of the EXP6 fluids. Unlike the
one Y model (1Y), multiple Y models have not been investigated in
detail yet and only a handful of results are
available~\cite{multipleY1,multipleY2}. To accomplish the goal,
i.e., to develop an analytic theory for the EXP6 fluid, we should first
know the properties of the 2Y fluids accurately and
in detail and this has been the motivation for the present study.

Since the virial coefficients provide the basic information on the
properties of the fluid and can be used in various theoretical
methods, we computed the first four virial coefficients of both
the parent EXP6 fluid and descending 2Y fluids. Furthermore, we
have also determined the vapor-liquid equilibrium (VLE) of the 2Y
fluid and located the critical point which is an important
information for setting the criteria of determining the 2Y fluid
associated with the EXP6 fluid.

\section{Basic definitions and computational details}

The EXP6 fluid is a collection of additive species (atoms, molecules, etc.)
interacting via the EXP6 potential (also referred to as a modified Buckingham
potential),
\begin{eqnarray}
u_{\ep}(r)=\left\{%
\begin{array}{ll}
    \infty , & \hbox{{\text for}\,\, $r < r_{\rm max}$\,,} \\
\displaystyle    \epsilon \left( \frac{6}{\alpha-6}
     \exp\left[\alpha (1-r/r_{\rm m})\right]- \frac{\alpha}{\alpha-6}(r_{\rm m}/r)^6\right),
     & \hbox{{\text for}\,\, $r > r_{\rm max}$\,,} \\
\end{array}%
\right.
\end{eqnarray}
where $r_{\rm max}$ and $\,r_{\rm m}\,$ is the location of the
potential maximum and minimum, respectively, and $\,\epsilon\,$ is
the depth of the minimum. Parameters $\,r_{\rm m}\,$ and
$\,\epsilon\,$ are used henceforth as the length and energy units;
dimensionless number density, $\rho^*$, temperature, $T^*$,
pressure, $P^*$, and internal energy, $U^*$, are thus defined as
$\rho^*=\rho r_{\rm m}^3$\,, $\,T^*=T\kB/\epsilon$, $P^*=Pr_{\rm
m}^3/\epsilon$, and $U^*=U/\epsilon$, respectively, where $\kB$ is
the Boltzmann constant.

The hard core 1Y potential is defined as
\begin{eqnarray}\label{Ypot}
u_{\YY}(r)=\left\{%
\begin{array}{ll}
    +\infty, & \hbox{{\rm for} $r<\sigma$,} \\
    (\sigma/r)\exp(-zr), & \hbox{{\rm for} $r \geqslant \sigma$,} \\
\end{array}%
\right.
\end{eqnarray}
where $z$ is the parameter governing the range of the interaction.

The 2Y potential is a model made up of two Yukawa tails without,
in general, any hard core,
\begin{equation}
u_{2Y}(r)=
 \epsilon_1\frac{\sigma}{r} \exp [-\kappa_1r] -
  \epsilon_2\frac{\sigma}{r} \exp [-\kappa_2r]\,,
\label{u2Y}
\end{equation}
where $\,\epsilon_1>0\,$ and $\,\epsilon_2>0\,$ are the strengths of the
repulsive and attractive contributions, respectively, while $\,\kappa_1^{-1}\,$
and $\,\kappa_2^{-1}\,$ are the measures of the range of the corresponding
tails.

The virial coefficients, $B_i$\,, of the virial expansion,
\begin{equation}
\frac{P}{\rho\kB T} = 1 + \sum_{i>1} B_i\rho^{i-1}   \,\,,
\end{equation}
were evaluated numerically up to $B_4$ over a wide range of
temperatures using the Mayer sampling~\cite{Mayer}. We have
recently examined another version of the virial expansion, the
perturbed expansion around a suitable reference system similarly
to the theories of liquids,
\begin{equation}
\frac{P}{\rho\kB T} = \left(\frac{P}{\rho\kB T}\right)_{\rf}
 + \sum_{i>1}\left[B_i(T) - B_{i,\rf}(T)\right]\rho^{i-1}  \,\,,
\end{equation}
where subscript ``ref'' refers to a reference system. To determine
the VLE envelope we used the common Gibbs ensemble with the total
number of particles $N=512$ and applied the long-range correction
in order to truncate the potential at $r_{\rm
c}=0.45\sqrt[3]{N/\rho^*}$\,.

\section{Results and discussion}

When the EXP6 model is applied to real fluids by adjusting its
parameter $\alpha$, its resulting values typically vary between 11
and 15. To keep contact with our previous
papers~\cite{Voertler,KNMT_2Y} we have chosen the bracketing
values, $\alpha=11.5$ and $14.5$. Parameters of the 2Y model
descending from these EXP6 models were determined in our previous
paper~\cite{KNMT_2Y} and are given in table~1.
\begin{table}[!h]
\centering \caption{The parameters of the 2Y potential function
defined by equation~(\ref{u2Y}) which are used to represent the
EXP6 fluid.} \label{2yp}
\vspace{1ex}
\footnotesize{
\begin{tabular}{llrrrrrrr}
\hline
 &$\alpha$ & $\epsilon_1/\epsilon$ & $\kappa_1 r_m$ & $\epsilon_2/\epsilon$ & $\kappa_2 r_m$
 && $\sigma/r_{\rm m}$ \\
 \hline
 SET I &   11.5    &   15026.86    &   9.4548  &   227.61   &  4.650    && 0.872    \\
 SET II    &   14.5    &   389565.31   &   13.344  &   148.24 &    4.514    && 0.892    \\
 \hline
\end{tabular}}
\vspace{5mm}
\end{table}

\begin{figure}[!h]
\begin{center}
\includegraphics[width=8cm]{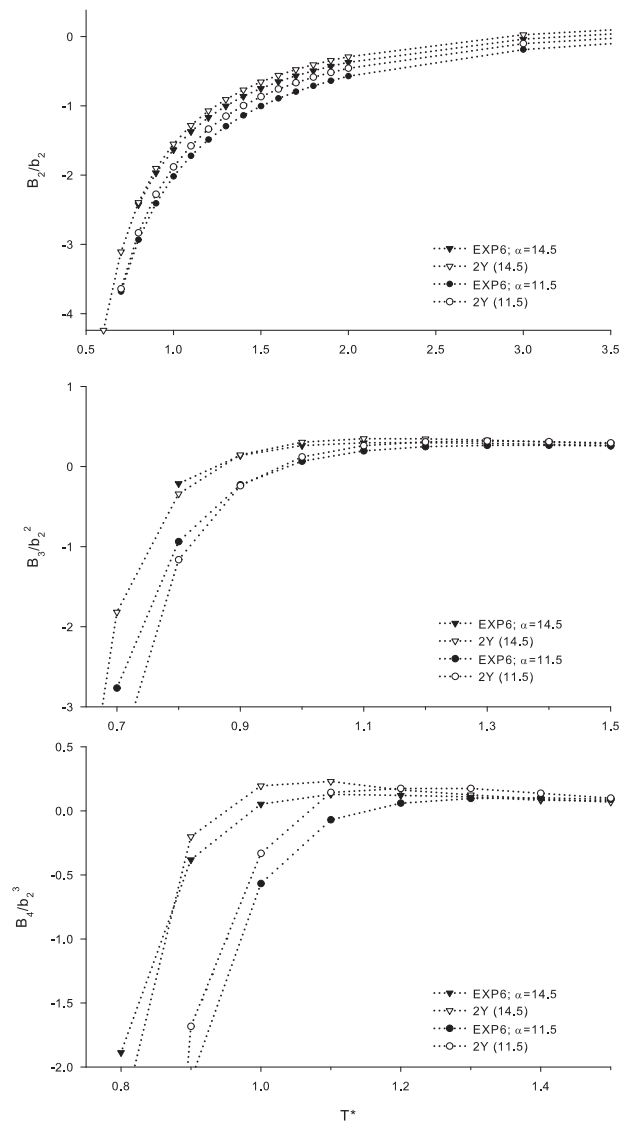}
\end{center}
\caption{Comparison of the virial coefficients $B_2$--$B_4$ of the considered models.
The dotted line has been drawn as a guide for eye.}
\label{fig}
\end{figure}
\begin{table}[ht]
\vspace{-7mm}
\centering \caption{Virial coefficients of the EXP6 potential with
$\alpha$=11.5; $b_2=({2}/{3})\pi r_{\rm m}^3$\,.}
\vspace{1ex}
\footnotesize{
\begin{tabular}{r r r r}
\hline
\multicolumn{1}{c}{$T^*$} & \multicolumn{1}{c}{$B_2/b_2$} & \multicolumn{1}{c}{$B_3/b_2^2$} & \multicolumn{1}{c}{$B_4/b_2^3$}\\
\hline
 50.00&     0.254 $\pm$      0.001&     0.049 $\pm$      0.001&     0.045 $\pm$      0.026\\
 40.00&     0.263 $\pm$      0.001&     0.057 $\pm$      0.001&     0.000 $\pm$      0.009\\
 30.00&     0.272 $\pm$      0.001&     0.066 $\pm$      0.001&     0.011 $\pm$      0.003\\
 20.00&     0.275 $\pm$      0.001&     0.080 $\pm$      0.001&     0.016 $\pm$      0.001\\
 15.00&     0.268 $\pm$      0.001&     0.090 $\pm$      0.001&     0.022 $\pm$      0.002\\
 14.00&     0.263 $\pm$      0.001&     0.092 $\pm$      0.001&     0.028 $\pm$      0.007\\
 13.00&     0.260 $\pm$      0.001&     0.094 $\pm$      0.001&     0.022 $\pm$      0.001\\
 12.00&     0.253 $\pm$      0.001&     0.097 $\pm$      0.001&     0.021 $\pm$      0.003\\
 11.00&     0.245 $\pm$      0.001&     0.100 $\pm$      0.001&     0.023 $\pm$      0.001\\
 10.00&     0.234 $\pm$      0.001&     0.103 $\pm$      0.001&     0.022 $\pm$      0.003\\
  9.00&     0.219 $\pm$      0.001&     0.106 $\pm$      0.001&     0.027 $\pm$      0.001\\
  8.00&     0.200 $\pm$      0.001&     0.111 $\pm$      0.001&     0.029 $\pm$      0.001\\
  7.00&     0.173 $\pm$      0.001&     0.115 $\pm$      0.001&     0.030 $\pm$      0.001\\
  6.00&     0.135 $\pm$      0.001&     0.120 $\pm$      0.001&     0.031 $\pm$      0.001\\
  5.00&     0.076 $\pm$      0.001&     0.127 $\pm$      0.001&     0.031 $\pm$      0.001\\
  4.00&    --0.018 $\pm$      0.001&     0.138 $\pm$      0.001&     0.031 $\pm$      0.001\\
  3.00&    --0.190 $\pm$      0.001&     0.158 $\pm$      0.001&     0.030 $\pm$      0.001\\
  2.00&    --0.573 $\pm$      0.001&     0.210 $\pm$      0.001&     0.046 $\pm$      0.002\\
  1.90&    --0.640 $\pm$      0.001&     0.219 $\pm$      0.001&     0.048 $\pm$      0.002\\
  1.80&    --0.714 $\pm$      0.001&     0.228 $\pm$      0.001&     0.057 $\pm$      0.002\\
  1.70&    --0.798 $\pm$      0.001&     0.239 $\pm$      0.001&     0.066 $\pm$      0.002\\
  1.60&    --0.895 $\pm$      0.001&     0.249 $\pm$      0.001&     0.076 $\pm$      0.002\\
  1.50&    --1.006 $\pm$      0.001&     0.258 $\pm$      0.001&     0.089 $\pm$      0.003\\
  1.40&    --1.139 $\pm$      0.001&     0.266 $\pm$      0.001&     0.101 $\pm$      0.004\\
  1.30&    --1.296 $\pm$      0.001&     0.264 $\pm$      0.001&     0.096 $\pm$      0.005\\
  1.20&    --1.489 $\pm$      0.001&     0.247 $\pm$      0.001&     0.059 $\pm$      0.006\\
  1.10&    --1.723 $\pm$      0.002&     0.196 $\pm$      0.001&    --0.071 $\pm$      0.009\\
  1.00&    --2.020 $\pm$      0.002&     0.066 $\pm$      0.002&    --0.569 $\pm$      0.014\\
  0.90&    --2.408 $\pm$      0.003&    --0.229 $\pm$      0.003&    --2.104 $\pm$      0.041\\
  0.80&    --2.934 $\pm$      0.003&    --0.939 $\pm$      0.004&    --7.230 $\pm$      0.230\\
  0.70&    --3.682 $\pm$      0.004&    --2.767 $\pm$      0.009&   --24.997 $\pm$      0.649\\
\hline
\end{tabular}}
\end{table}
\begin{table}[!h]
\vspace{-3mm}
\centering \caption{Virial coefficients of the EXP6 potential with
$\alpha=14.5$; $b_2=({2}/{3})\pi r_{\rm m}^3$\,.}
\vspace{1ex}
\footnotesize{
\begin{tabular}{r r r r}
\hline
\multicolumn{1}{c}{$T^*$} & \multicolumn{1}{c}{$B_2/b_2$} & \multicolumn{1}{c}{$B_3/b_2^2$} & \multicolumn{1}{c}{$B_4/b_2^3$}\\
\hline
 50.00&     0.340 $\pm$      0.001&     0.083 $\pm$      0.001&     0.016 $\pm$      0.002\\
 40.00&     0.351 $\pm$      0.001&     0.091 $\pm$      0.001&     0.023 $\pm$      0.004\\
 30.00&     0.360 $\pm$      0.001&     0.103 $\pm$      0.001&     0.023 $\pm$      0.003\\
 20.00&     0.364 $\pm$      0.001&     0.116 $\pm$      0.001&     0.030 $\pm$      0.002\\
 15.00&     0.358 $\pm$      0.001&     0.126 $\pm$      0.001&     0.039 $\pm$      0.007\\
 14.00&     0.355 $\pm$      0.001&     0.129 $\pm$      0.001&     0.036 $\pm$      0.002\\
 13.00&     0.351 $\pm$      0.001&     0.131 $\pm$      0.001&     0.034 $\pm$      0.002\\
 12.00&     0.347 $\pm$      0.001&     0.135 $\pm$      0.001&     0.035 $\pm$      0.001\\
 11.00&     0.340 $\pm$      0.001&     0.135 $\pm$      0.001&     0.033 $\pm$      0.007\\
 10.00&     0.330 $\pm$      0.001&     0.138 $\pm$      0.001&     0.042 $\pm$      0.001\\
  9.00&     0.318 $\pm$      0.001&     0.142 $\pm$      0.001&     0.044 $\pm$      0.002\\
  8.00&     0.301 $\pm$      0.001&     0.144 $\pm$      0.001&     0.041 $\pm$      0.002\\
  7.00&     0.278 $\pm$      0.001&     0.147 $\pm$      0.001&     0.045 $\pm$      0.001\\
  6.00&     0.245 $\pm$      0.001&     0.150 $\pm$      0.001&     0.049 $\pm$      0.001\\
  5.00&     0.195 $\pm$      0.001&     0.154 $\pm$      0.001&     0.048 $\pm$      0.001\\
  4.00&     0.112 $\pm$      0.001&     0.160 $\pm$      0.001&     0.047 $\pm$      0.001\\
  3.00&    --0.039 $\pm$      0.001&     0.171 $\pm$      0.001&     0.043 $\pm$      0.001\\
  2.00&    --0.373 $\pm$      0.001&     0.209 $\pm$      0.001&     0.039 $\pm$      0.001\\
  1.90&    --0.429 $\pm$      0.001&     0.216 $\pm$      0.001&     0.041 $\pm$      0.001\\
  1.80&    --0.495 $\pm$      0.001&     0.225 $\pm$      0.001&     0.045 $\pm$      0.001\\
  1.70&    --0.568 $\pm$      0.001&     0.236 $\pm$      0.001&     0.051 $\pm$      0.002\\
  1.60&    --0.652 $\pm$      0.001&     0.249 $\pm$      0.001&     0.056 $\pm$      0.002\\
  1.50&    --0.751 $\pm$      0.001&     0.263 $\pm$      0.001&     0.078 $\pm$      0.006\\
  1.40&    --0.863 $\pm$      0.001&     0.276 $\pm$      0.001&     0.082 $\pm$      0.003\\
  1.30&    --1.003 $\pm$      0.001&     0.289 $\pm$      0.001&     0.108 $\pm$      0.004\\
  1.20&    --1.169 $\pm$      0.001&     0.297 $\pm$      0.001&     0.120 $\pm$      0.005\\
  1.10&    --1.373 $\pm$      0.002&     0.296 $\pm$      0.001&     0.128 $\pm$      0.007\\
  1.00&    --1.635 $\pm$      0.002&     0.262 $\pm$      0.002&     0.052 $\pm$      0.010\\
  0.90&    --1.969 $\pm$      0.002&     0.137 $\pm$      0.002&    --0.382 $\pm$      0.019\\
  0.80&    --2.427 $\pm$      0.002&    --0.213 $\pm$      0.002&    --1.888 $\pm$      0.160\\
\hline
\end{tabular}}
\vspace{-5mm}
\end{table}
\begin{table}[!h]
\vspace{-7mm}
\centering \caption{Virial coefficients of the 2-Yukawa potential
mimicking the EXP6 potential with $\alpha=11.5$; $b_2=({2}/{3})\pi
\sigma^3$.}
\footnotesize{
\begin{tabular}{r r r r}
\hline
\multicolumn{1}{c}{$T^*$} & \multicolumn{1}{c}{$B_2/b_2$} & \multicolumn{1}{c}{$B_3/b_2^2$} & \multicolumn{1}{c}{$B_4/b_2^3$}\\
\hline
 50.00&     0.268 $\pm$      0.001&     0.052 $\pm$      0.001&     0.007 $\pm$      0.001\\
 40.00&     0.278 $\pm$      0.001&     0.059 $\pm$      0.001&     0.010 $\pm$      0.001\\
 30.00&     0.290 $\pm$      0.001&     0.069 $\pm$      0.001&     0.013 $\pm$      0.001\\
 20.00&     0.297 $\pm$      0.001&     0.081 $\pm$      0.001&     0.017 $\pm$      0.001\\
 15.00&     0.293 $\pm$      0.001&     0.092 $\pm$      0.001&     0.019 $\pm$      0.003\\
 14.00&     0.291 $\pm$      0.001&     0.094 $\pm$      0.001&     0.022 $\pm$      0.001\\
 13.00&     0.287 $\pm$      0.001&     0.097 $\pm$      0.001&     0.023 $\pm$      0.001\\
 12.00&     0.282 $\pm$      0.001&     0.100 $\pm$      0.001&     0.015 $\pm$      0.013\\
 11.00&     0.276 $\pm$      0.001&     0.103 $\pm$      0.001&     0.024 $\pm$      0.001\\
 10.00&     0.268 $\pm$      0.001&     0.106 $\pm$      0.001&     0.025 $\pm$      0.001\\
  9.00&     0.257 $\pm$      0.001&     0.108 $\pm$      0.001&     0.028 $\pm$      0.001\\
  8.00&     0.241 $\pm$      0.001&     0.112 $\pm$      0.001&     0.030 $\pm$      0.001\\
  7.00&     0.218 $\pm$      0.001&     0.116 $\pm$      0.001&     0.031 $\pm$      0.001\\
  6.00&     0.186 $\pm$      0.001&     0.120 $\pm$      0.001&     0.032 $\pm$      0.001\\
  5.00&     0.134 $\pm$      0.001&     0.127 $\pm$      0.001&     0.032 $\pm$      0.001\\
  4.00&     0.051 $\pm$      0.001&     0.137 $\pm$      0.001&     0.032 $\pm$      0.001\\
  3.00&    --0.104 $\pm$      0.001&     0.159 $\pm$      0.001&     0.028 $\pm$      0.001\\
  2.00&    --0.458 $\pm$      0.001&     0.223 $\pm$      0.001&     0.038 $\pm$      0.001\\
  1.90&    --0.520 $\pm$      0.001&     0.234 $\pm$      0.001&     0.042 $\pm$      0.001\\
  1.80&    --0.588 $\pm$      0.001&     0.249 $\pm$      0.001&     0.051 $\pm$      0.002\\
  1.70&    --0.669 $\pm$      0.001&     0.262 $\pm$      0.001&     0.068 $\pm$      0.005\\
  1.60&    --0.760 $\pm$      0.001&     0.278 $\pm$      0.001&     0.087 $\pm$      0.003\\
  1.50&    --0.870 $\pm$      0.001&     0.295 $\pm$      0.001&     0.099 $\pm$      0.003\\
  1.40&    --0.998 $\pm$      0.001&     0.310 $\pm$      0.001&     0.136 $\pm$      0.004\\
  1.30&    --1.151 $\pm$      0.001&     0.319 $\pm$      0.001&     0.174 $\pm$      0.006\\
  1.20&    --1.338 $\pm$      0.002&     0.307 $\pm$      0.001&     0.174 $\pm$      0.008\\
  1.10&    --1.580 $\pm$      0.002&     0.260 $\pm$      0.002&     0.143 $\pm$      0.012\\
  1.00&    --1.883 $\pm$      0.002&     0.119 $\pm$      0.002&    --0.333 $\pm$      0.021\\
  0.90&    --2.278 $\pm$      0.002&    --0.240 $\pm$      0.003&    --1.683 $\pm$      0.308\\
  0.80&    --2.834 $\pm$      0.003&    --1.165 $\pm$      0.004&    --8.430 $\pm$      0.150\\
\hline
\end{tabular}}
\vspace{-3mm}
\end{table}
\begin{table}[!h]
\centering \caption{Virial coefficients of the 2-Yukawa potential
mimicking the EXP6 potential with $\alpha=14.5$;
$b_2=({2}/{3})\pi\sigma^3$.}
\footnotesize{
\begin{tabular}{r r r r}
\hline
\multicolumn{1}{c}{$T^*$}&\multicolumn{1}{c}{$B_2/b_2$}&\multicolumn{1}{c}{$B_3/b_2^2$}&\multicolumn{1}{c}{$B_4/b_2^3$}\\
\hline
 50.00&     0.355 $\pm$      0.001&     0.087 $\pm$      0.001&     0.015 $\pm$      0.002\\
 40.00&     0.363 $\pm$      0.001&     0.095 $\pm$      0.001&     0.027 $\pm$      0.006\\
 30.00&     0.376 $\pm$      0.001&     0.106 $\pm$      0.001&     0.021 $\pm$      0.002\\
 20.00&     0.383 $\pm$      0.001&     0.120 $\pm$      0.001&     0.031 $\pm$      0.001\\
 15.00&     0.380 $\pm$      0.001&     0.130 $\pm$      0.001&     0.028 $\pm$      0.006\\
 14.00&     0.378 $\pm$      0.001&     0.131 $\pm$      0.001&     0.036 $\pm$      0.002\\
 13.00&     0.374 $\pm$      0.001&     0.134 $\pm$      0.001&     0.035 $\pm$      0.004\\
 12.00&     0.371 $\pm$      0.001&     0.135 $\pm$      0.001&     0.039 $\pm$      0.001\\
 11.00&     0.365 $\pm$      0.001&     0.137 $\pm$      0.001&     0.046 $\pm$      0.005\\
 10.00&     0.359 $\pm$      0.001&     0.140 $\pm$      0.001&     0.042 $\pm$      0.001\\
  9.00&     0.348 $\pm$      0.001&     0.143 $\pm$      0.001&     0.042 $\pm$      0.001\\
  8.00&     0.334 $\pm$      0.001&     0.145 $\pm$      0.001&     0.045 $\pm$      0.001\\
  7.00&     0.314 $\pm$      0.001&     0.147 $\pm$      0.001&     0.046 $\pm$      0.001\\
  6.00&     0.284 $\pm$      0.001&     0.150 $\pm$      0.001&     0.049 $\pm$      0.001\\
  5.00&     0.239 $\pm$      0.001&     0.153 $\pm$      0.001&     0.049 $\pm$      0.001\\
  4.00&     0.163 $\pm$      0.001&     0.158 $\pm$      0.001&     0.047 $\pm$      0.001\\
  3.00&     0.026 $\pm$      0.001&     0.168 $\pm$      0.001&     0.041 $\pm$      0.001\\
  2.00&    --0.292 $\pm$      0.001&     0.214 $\pm$      0.001&     0.033 $\pm$      0.001\\
  1.90&    --0.345 $\pm$      0.001&     0.225 $\pm$      0.001&     0.031 $\pm$      0.001\\
  1.80&    --0.407 $\pm$      0.001&     0.236 $\pm$      0.001&     0.038 $\pm$      0.002\\
  1.70&    --0.478 $\pm$      0.001&     0.250 $\pm$      0.001&     0.043 $\pm$      0.002\\
  1.60&    --0.560 $\pm$      0.001&     0.268 $\pm$      0.001&     0.050 $\pm$      0.002\\
  1.50&    --0.655 $\pm$      0.001&     0.286 $\pm$      0.001&     0.068 $\pm$      0.002\\
  1.40&    --0.771 $\pm$      0.001&     0.306 $\pm$      0.001&     0.091 $\pm$      0.004\\
  1.30&    --0.909 $\pm$      0.001&     0.329 $\pm$      0.001&     0.124 $\pm$      0.005\\
  1.20&    --1.074 $\pm$      0.001&     0.345 $\pm$      0.001&     0.162 $\pm$      0.006\\
  1.10&    --1.283 $\pm$      0.001&     0.348 $\pm$      0.001&     0.229 $\pm$      0.010\\
  1.00&    --1.553 $\pm$      0.001&     0.304 $\pm$      0.002&     0.195 $\pm$      0.015\\
  0.90&    --1.904 $\pm$      0.002&     0.146 $\pm$      0.002&    --0.202 $\pm$      0.023\\
  0.80&    --2.397 $\pm$      0.002&    --0.342 $\pm$      0.003&    --2.433 $\pm$      0.056\\
  0.70&    --3.108 $\pm$      0.003&    --1.818 $\pm$      0.007&   --14.300 $\pm$      0.330\\
  0.60&    --4.239 $\pm$      0.004&    --6.815 $\pm$      0.025&   --86.259 $\pm$      2.362\\
\hline
\end{tabular}}
\vspace{-12mm}
\end{table}
Numerical values of the virial coefficients of all four fluids
considered are given in tables~2 through~5 and are also compared
in figure~\ref{fig}. Examination of the tables/figure shows that the
coefficients of the EXP6 and 2Y fluids are very similar. This is a
consequence of the fact that the repulsive parts of the EXP6 and
2Y models have been used to fit the corresponding 2nd virial
coefficients~\cite{KNMT_2Y}. In other words, the similarity of all
the virial coefficients means that the long-range part of the
models does not affect them to any important extent.

\begin{table}[!h]
\centering
\caption{Vapor-liquid equilibrium data of the 2Y fluid mimicking the EXP6 potential with $\alpha=11.5$\,.}
\footnotesize{
\begin{tabular}{r r r}
\hline
\multicolumn{1}{c}{$T^*$}&\multicolumn{1}{c}{$\rho^*_v$}&\multicolumn{1}{c}{$\rho^*_l$}\\
\hline
 0.700 & 0.0073 $\pm$ 0.0031 & 1.1798 $\pm$ 0.0130\\
 0.800 & 0.0194 $\pm$ 0.0060 & 1.1097 $\pm$ 0.0162\\
 0.900 & 0.0438 $\pm$ 0.0086 & 1.0331 $\pm$ 0.0177\\
 1.000 & 0.0886 $\pm$ 0.0159 & 0.9327 $\pm$ 0.0294\\
 1.030 & 0.1107 $\pm$ 0.0183 & 0.9012 $\pm$ 0.0352\\
 1.050 & 0.1333 $\pm$ 0.0253 & 0.8748 $\pm$ 0.0387\\
 1.070 & 0.1567 $\pm$ 0.0288 & 0.8540 $\pm$ 0.0353\\
 1.100 & 0.1930 $\pm$ 0.0439 & 0.7805 $\pm$ 0.0610\\
 1.120 & 0.2324 $\pm$ 0.0282 & 0.7203 $\pm$ 0.0577\\
 1.140 & 0.2732 $\pm$ 0.0502 & 0.6241 $\pm$ 0.0717\\
\hline
\end{tabular}}
\end{table}
\begin{table}[ht]
\centering
\caption{Vapor-liquid equilibrium data of the 2Y fluid mimicking the EXP6 potential with $\alpha=14.5$\,.}
\footnotesize{
\begin{tabular}{r r r}
\hline
\multicolumn{1}{c}{$T^*$}&\multicolumn{1}{c}{$\rho^*_v$}&\multicolumn{1}{c}{$\rho^*_l$}\\
\hline
 0.600 & 0.0054 $\pm$ 0.0028 & 1.1571 $\pm$ 0.0113\\
 0.700 & 0.0172 $\pm$ 0.0049 & 1.0863 $\pm$ 0.0152\\
 0.800 & 0.0454 $\pm$ 0.0105 & 0.9983 $\pm$ 0.0193\\
 0.850 & 0.0688 $\pm$ 0.0132 & 0.9437 $\pm$ 0.0219\\
 0.900 & 0.1055 $\pm$ 0.0167 & 0.8799 $\pm$ 0.0283\\
 0.930 & 0.1394 $\pm$ 0.0251 & 0.8287 $\pm$ 0.0459\\
 0.950 & 0.1533 $\pm$ 0.0256 & 0.7852 $\pm$ 0.0603\\
 0.970 & 0.1948 $\pm$ 0.0257 & 0.7512 $\pm$ 0.0462\\
 0.975 & 0.2072 $\pm$ 0.0319 & 0.7287 $\pm$ 0.0576\\
 0.980 & 0.2157 $\pm$ 0.0341 & 0.7078 $\pm$ 0.0584\\
 0.985 & 0.2211 $\pm$ 0.0348 & 0.6920 $\pm$ 0.0604\\
 0.990 & 0.2320 $\pm$ 0.0238 & 0.6677 $\pm$ 0.0527\\
 0.995 & 0.2474 $\pm$ 0.0548 & 0.6012 $\pm$ 0.0720\\
\hline
\end{tabular}}
\vspace{-5mm}
\end{table}
\begin{table}[ht]
\centering
\caption{Critical properties of the 2Y fluids determined from the vapor-liquid coexistence data. Numbers in parentheses give results of the perturbed virial expansion (first row) and the virial expansion (second row) of the second order.}
\vspace{2ex}
\footnotesize{
\begin{tabular}{l r rr rr}
\hline
\multicolumn{1}{c}{Model}
&\multicolumn{1}{c}{$\rho^*$}&\multicolumn{1}{c}{$T^*$}\\
\hline
2Y--11.5 & 0.452 &  1.171\\
        &  &  (1.18)\\
        &  &  (1.43)\\
2Y--14.5 & 0.295 &  1.00 \\
        &  &  (1.05)\\
        &  &  (1.24)\\
\hline
\end{tabular}}
\end{table}

The VLE results for the equilibrium densities of the 2Y fluids are
given in tables~6 and~7. With these data, the critical point was
determined using the rectilinear rule and the common analytic
parametrization expression
\begin{equation}
\rho_{\rm l}-\rho_{\rm
v}=B_0\left|t\right|^\beta+B_1\left|t\right|^{\beta+\Delta_1},
\end{equation}
and
\begin{equation}
\frac{\rho_{\rm l}+\rho_{\rm v}}{2}=\rho_{\rm
c}+C_1\left|t\right|^\psi+C_2\left|t\right|+C_3\left|t\right|^{\psi+\Delta_1},
\end{equation}
where $t=1-T/T_{\rm C}$ and $B_i$\,, $C_i$\,, $\beta$ and $\psi$
are parameters to be fitted to the simulation data; $\Delta_1=0.5$
for the vapor-liquid equilibria~\cite{Jackson}. The results are
given in table~8. In this table we also give an estimate of the
critical temperature obtained using the common virial expansion
and the perturbed virial expansion of the 2nd
order~\cite{NScrit,KNcrit}; in the perturbed expansion we used the
fluid of hard spheres of diameter $\sigma$ as the reference. As we
see, the perturbed method provides a very good estimate which
further confirms its superiority over the common virial expansion.

\section{Conclusions}

In this paper we have presented results for the first four virial coefficients of the EXP6 fluid and the associated 2-Yukawa fluids, and vapor-liquid equilibria in the latter models. These data are necessary for a subsequent development of a theory for the 2-Yukawa fluids which further provides an alternative to hard core (van der Waals type) equations of state. In addition to the determination of the critical point from the vapor-liquid coexistence data, we have used the computed virial coefficients to determine the critical temperature from the virial expansions. The obtained results further confirm the recently reported applicability and accuracy of the perturbed expansion.

\section*{Acknowledgement}

This work was supported by the Grant Agency of the Academy of
Sciences of the Czech Republic (Grant No.~IAA400720710), by the
Grant Agency of the J.E.~Purkinje University (Grant
No.~53223--15--0010--01), and the Czech-Ukrainian Bilateral
Cooperative Program.

\newpage
\ukrainianpart

\title{Віріальні коефіцієнти і фазова рівновага пара-рідина у EXP6 та 2-Юкава плинах}
\author
{Я. Крейчі\refaddr{label1},
        І. Незбеда\refaddr{label1,label2},
        Р. Мельник\refaddr{label3},
        А. Трохимчук\refaddr{label3}
}
\addresses{
\addr{label1} Факультет науки, Університет Я.Е. Пуркінйє, 400 96 Усті над Лабем,  Чеська Республіка
\addr{label2} Інститут фундаментальних основ хімічних процесів, Академія наук, 165 02 Прага 6, \\Чеська Республіка
\addr{label3}
Інститут фізики конденсованих систем НАН України,
79011 Львів, вул. Свєнціцького, 1}

\makeukrtitle

\begin{abstract}
\tolerance=3000%
Віріальні коефіцієнти від $B_2$ до $B_4$ і фазова рівновага пара-рідина у EXP6 та 2-Юкава (2Y) плинах розраховані, відповідно, з допомогою чисельного інтегрування та на основі комп'ютерного експерименту з використанням ансамблю Гібса. Вибрані 2Y модельні системи нещодавно були запропоновані  як  базисні для EXP6 плинів, що розглядаються.%
\keywords EXP6 плин, 2-Юкава плин, віріальні коефіцієнти, фазова рівновага пара-рідина, критична точка
\end{abstract}

\end{document}